\documentclass[journal, 10pt]{IEEEtran}

\usepackage{multirow}
\usepackage[font=scriptsize,caption=false,labelsep=space]{subfig}
\usepackage{mathrsfs}
\usepackage{graphicx,float,wrapfig,epstopdf,amsmath}
\usepackage[]{algorithmicx}
\usepackage{algpseudocode,algorithm}
\usepackage{balance}
\usepackage{wrapfig}
\usepackage{enumitem}

\usepackage{mathtools,lipsum}
\usepackage{amsmath}
\usepackage{amssymb}
\usepackage{amsthm}
\usepackage{graphicx}
\usepackage{epstopdf}
\usepackage{times}
\usepackage{textcomp,cite}

\usepackage{url}
\usepackage{hyperref}

\usepackage{multirow}
\usepackage{threeparttable}
\graphicspath{{./figures/}}

\DeclareMathOperator*{\maximize}{max} 
\DeclareMathOperator*{\minimize}{minimize} 
\DeclareMathOperator*{\subjectto}{s.\hspace{1pt} t.:\hspace{3pt}} 

\usepackage[table]{xcolor}
\definecolor{intnull}{RGB}{213,229,255}
\definecolor{inteins}{RGB}{128,179,255}
\definecolor{color1}{RGB}{199,209,232}
\definecolor{color2}{RGB}{230,231,233}

\begin{document}

	\title{ Federated Multi-Task Learning \\ for THz Wideband Channel and DoA Estimation
	}

	\author{\IEEEauthorblockN{Ahmet M. Elbir, \textit{Senior Member, IEEE,}
			Wei Shi, \textit{Member, IEEE,}
			Kumar Vijay Mishra, \textit{Senior Member, IEEE,}
			and  Symeon Chatzinotas, \textit{Senior Member, IEEE}
		}
		\thanks{This work was supported in part by the Natural Sciences and Engineering Research Council of Canada (NSERC), Ericsson Canada, and the ERC Project AGNOSTIC.}
		\thanks{A. M. Elbir is with 
			Interdisciplinary Centre for Security, Reliability and Trust (SnT) at the University of Luxembourg, Luxembourg (e-mail: ahmetmelbir@gmail.com).} 
		\thanks{W. Shi is with the	School of Information Technology, Carleton
			University, Ottawa, Canada (e-mail: wei.shi@carleton.ca).}
		\thanks{K. V. Mishra is with the United States Army Research Laboratory, Adelphi, MD 20783 USA (e-mail: kvm@ieee.org).}
		
		\thanks{S. Chatzinotas is with the SnT at the University of Luxembourg, Luxembourg (e-mail: symeon.chatzinotas@uni.lu). }
	}
	\maketitle
	
	\begin{abstract}
		This paper addresses two major challenges in terahertz (THz) channel estimation: the beam-split phenomenon, i.e., beam misalignment because of frequency-independent analog beamformers, and computational complexity because of the usage of ultra-massive number of antennas to compensate propagation losses. Data-driven techniques are known to mitigate the complexity of this problem but usually require the transmission of the datasets from the users to a central server entailing huge communication overhead. In this work, we introduce a federated multi-task learning (FMTL), wherein the users transmit only the model parameters instead of the whole dataset, for THz channel and user direction-of-arrival (DoA) estimation to improve the communications-efficiency. We first propose a novel beamspace support alignment technique for channel estimation with beam-split correction. Then, the channel and DoA information are used as labels to train an FMTL model. By exploiting the sparsity of the THz channel, the proposed approach is implemented with fewer pilot signals than the traditional techniques. Compared to the previous works, our FMTL approach provides higher channel estimation accuracy as well as approximately 25 (32) times lower model (channel) training overhead, respectively.

	\end{abstract}
	%
	

	\section{Introduction}
	\label{sec:Introduciton}
	
	\textcolor{black}{Compared to millimeter-wave (mm-Wave) frequencies, propagation at terahertz (THz) band is beset with several challenges such as spreading loss, molecular absorption, and more severe attenuation leading to a shorter range. To compensate these loses via beamforming gain, ultra-massive (UM) antenna arrays are used~\cite{ummimoHBThzSVModel}.} The THz channels are extremely sparse and modeled as line-of-sight (LoS)-dominant and non-LoS (NLoS)-assisted models~\cite{ummimoTareqOverview,ummimoHBThzSVModel,thz_beamSplit}. On the other hand, both LoS and NLoS paths are significant in the mm-Wave channel. Furthermore, the use of frequency-independent analog beamformers, \textit{beam-split} phenomenon occurs in THz-bands, i.e., the generated beams split into different physical directions (see, e.g., \cite[Fig.11]{elbir2021JointRadarComm} and \cite[Fig.4]{trueTimeDelayPrecoding_THz_Dai2022Mar}  for illustration) at each subcarrier because of ultra-wide bandwidth and large number of antennas~\cite{thz_beamSplit}.
	
	With the aforementioned unique peculiarities, THz channel estimation is a challenging problem. In prior works, the beam-split effect remains relatively unexamined~\cite{channelEstThz,thz_CE_GAN_Balevi2021Jan}. Specifically, beam-split causes enough deviations in the spatial channel directions at different subcarriers, producing separation in the angular domain~\cite{channelEstThz_beamSplit}. At mm-Wave, \textit{beam-squint} is broadly used to describe the same phenomenon~\cite{elbir2021JointRadarComm}). While~\cite{channelEstThz2} and \cite{trueTimeDelayPrecoding_THz_Dai2022Mar} employ time-delay networks as an additional hardware component to mitigate the beam-split, signal processing techniques have also been considered~\cite{elbir2021JointRadarComm}. For instance, \cite{channelEstThz_beamSplit} developed a beam-split pattern detection (BSPD) technique to recover the channel support among subcarriers and construct a one-to-one match for the physical channel direction. 
	
	Besides the above model-based techniques, data-driven model-free approaches, e.g., machine learning (ML), have also been suggested for THz channel estimation~\cite{thzCE_kernel,thzCE_CNN,thz_dl_Gao2022May}. In particular, ML-based models, e.g., deep convolutional neural network (DCNN)~\cite{thzCE_CNN}, generative adversarial network (GAN)~\cite{thz_CE_GAN_Balevi2021Jan}, and deep kernel learning (DKL)~\cite{thzCE_kernel}, have been introduced to lower the computational complexity arising from UM multiple-input multiple-output (MIMO) architectures. 
	However, all of these works consider a narrowband scenario, which does not exploit the key drivers of wide bandwidths in the THz band. \textcolor{black}{The unfolded deep neural network (UDNN) architecture used in~\cite{thz_dl_Gao2022May} for mm-Wave channel estimation in the presence of beam-squint assumes subcarrier-independent steering vectors to construct the wideband channel. Hence, its performance is limited and, indeed, no performance improvement was observed by enhancing the grid resolution.}
	
	{\color{black}The ML-based techniques involve the transmission of the datasets to a central parameter server (PS), entailing huge communications overhead. This is significant at THz band as compared to mm-Wave because of multi-fold increase in the number of antennas. In order to reduce the amount of transmitted data, federated learning (FL)-based channel estimation is proposed} in~\cite{elbir2020_FL_CE}, wherein the CNN parameters are computed at the users and, in place of the datasets, these parameters are sent to the PS for model aggregation. Compared to centralized learning (CL) models, FL-based approaches are reported to have approximately $20$-$30$ times lower the communication overhead~\cite{elbir2020_FL_CE}. \textcolor{black}{However, \cite{elbir2020_FL_CE} only considers mm-Wave channel estimation and the effect of beam-split is not taken into account.}
	
\begin{figure}[t]
		\centering
		{\includegraphics[draft=false,width=.8\columnwidth]{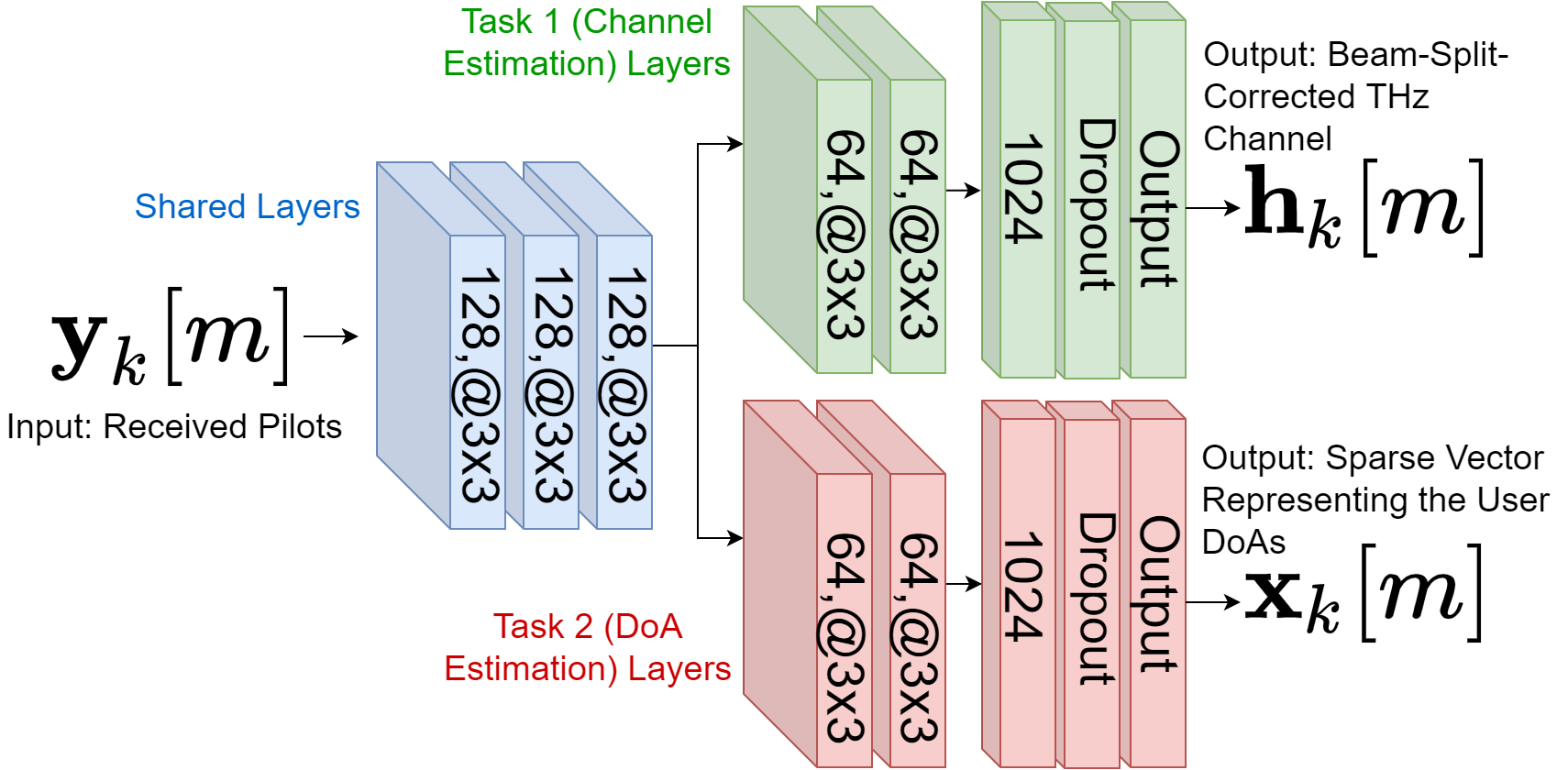} } 
		\caption{An Illustration of FMTL  model architecture.
		}
		\label{fig_Model}
	\end{figure}

	In this work, we propose a federated multi-task learning (FMTL) approach to jointly estimate wideband THz channel and user direction-of-arrivals (DoAs) in the presence of beam-split. 	Our main contributions of this work are:
	\begin{enumerate}[wide]
		\item We first introduce a novel model-based approach to suppress the effect of beam-split for THz wideband channel estimation, called beamspace support alignment (BSA) technique, in which the received signal is converted to beamspace. Here, the amount of angular deviation because of beam-split is determined based on the mismatch between the central and subcarrier frequencies. Then, the beamspace spectra of subcarriers are shifted and combined to generate a single spectrum such that the channel supports are aligned.
		\item By selecting the beam-split-corrected channel and DoAs as labels,  we then propose an FMTL-based approach by training a learning model (Fig.~\ref{fig_Model}) on local datasets to perform joint beam-split-corrected channel and DoA estimation.
		\item We also investigate the channel estimation performance and communications-efficiency of FMTL and CL via numerical simulations, which demonstrate approximately $25$ ($32$) times less overhead for model (channel) training while maintaining satisfactory channel estimation performance in the presence of beam-split.
	\end{enumerate}

	
	
	%

	\section{Signal Model and Problem Formulation}
	\label{sec:probForm}
	Consider a wideband ultra-massive MIMO architecture, wherein the base station (BS) is equipped with $N_\mathrm{T}$ antennas and $N_\mathrm{RF}$ radio-frequency (RF) chains to serve $K$ single-antenna users. In the downlink, the BS employs frequency-dependent baseband beamformer $\mathbf{F}_\mathrm{BB}[m]\in \mathbb{C}^{K\times K}$  to process the transmitted signals $\mathbf{s}[m] = [s_1[m],\dots,s_K[m]]^\textsf{T}$ ($m\in \mathcal{M} = \{1,\dots, M\}$). Then, frequency-independent analog beamformers $\mathbf{F}_\mathrm{RF}\in \mathbb{C}^{N_\mathrm{T}\times N_\mathrm{RF}}$ ($N_\mathrm{RF}=K<{N}_\mathrm{T}$)  are used to steer the generated beams toward users. Therefore, $\mathbf{F}_\mathrm{RF}$ has unit-modulus constraint, i.e., $|[\mathbf{F}_\mathrm{RF}]_{i,j}| = \frac{1}{\sqrt{N_\mathrm{T}}}$ as $i = 1,\dots, N_\mathrm{RF}$ and $j = 1,\dots, N_\mathrm{T}$.	  Let us define the transmitted signal as  $\mathbf{z}[m] = \mathbf{F}_\mathrm{RF}\mathbf{F}_\mathrm{BB}[m]\mathbf{s}[m]$, then 	the received signal at the $k$th user for the $m$th subcarrier is 
	\begin{align}
	\label{receivedSignal}
	{y}_{k}[m] = \mathbf{h}_{k}^\textsf{T}[m]\mathbf{z}[m]  + {n}_k[m],
	\end{align}
	where ${n}_k[m]\in \mathbb{C}$ denotes the additive white Gaussian noise (AWGN) vector with $\mathbf{n}_k[m] \sim \mathcal{CN}({0},\sigma_n^2)$.

	In THz transmission, the wireless channel is modeled as the superposition of a single LoS path and the contribution of a few NLoS paths, which are small due to large reflection loses, scattering and refraction~\cite{ummimoTareqOverview,ummimoHBThzSVModel,thz_beamSplit}. Then, the $N_\mathrm{T}\times 1$ THz channel vector corresponding to the $k$th user at $m$th subcarrier is 
	\begin{align}
	\label{channelModel}
	\hspace{-3pt}\mathbf{h}_k[m] \hspace{-3pt} =  \hspace{-3pt}
	\gamma\big( \hspace{-1pt}{\alpha_{k}^{m,1} \mathbf{a}(\theta_{k,m,1}) }
	\hspace{-1pt}+  \hspace{-5pt}{\sum_{l =2}^{L}  \hspace{-3pt}\alpha_{k}^{m,l} \mathbf{a}(\theta_{k,m,l})} \hspace{-1pt} \big) e^{-j2\pi\tau_{k,l} f_m },
	\end{align}
	where $\gamma = \sqrt{\frac{N_\mathrm{T}}{L}}$, $\alpha_{k}^{m,l}\in\mathbb{C}$ is the complex path gain, $L$ is the total number of paths, and $\tau_{k,l}$ is the time delay of the $l$th path. $f_m = f_c + \frac{B}{M}(m - 1 - \frac{M-1}{2}) $ is the $m$th subcarrier frequency with $f_c$ and $B$ being the carrier frequency and the bandwidth, respectively.  The steering vector corresponding to the spatial channel DoA $\theta_{k,m,l}$ defined for a uniform linear array is $N_T \times 1$ vector $	\mathbf{a}(\theta_{k,m,l}) \hspace{-3pt}= \hspace{-3pt} \frac{1}{\sqrt{N_\mathrm{T}}} [1, e^{-j \pi  \theta_{k,m,l} },\dots, e^{-j \pi (N_\mathrm{T}-1) \theta_{k,m,l} }]^\textsf{T},$
	where
	\begin{align}
	\label{physical_spatial_directions}
	\theta_{k,m,l} =  \frac{2 f_m}{c_0} d \vartheta_{k,l} =  \frac{f_m}{f_c} \vartheta_{k,l},
	\end{align}
	denotes the spatial DoA with $\vartheta = \sin \tilde{\vartheta}_{k,l} $ being the physical DoA for the $l$-th path with $\tilde{\vartheta}_{k,l} \in [-\frac{\pi}{2}, \frac{\pi}{2}] $,	$c_0$ is the speed of light, and $d = \frac{c_0}{2f_c}$ is the half-wavelength antenna spacing.
	
	Note that $\theta_{k,m,l} \approx \vartheta_{k,l}$ when $f_m \approx f_c$.
	This observation allows us to employ frequency-independent analog beamformers (i.e., $\mathbf{F}_\mathrm{RF}$) for $m\in\mathcal{M}$ in conventional mm-Wave systems. However, beam-split implies that with wider  bandwidth $|f_m - f_c|$, physical DoA $\vartheta_{k,l}$ deviate from the spatial DoA $\theta_{k,m,l}$. 
	
	Our goal is to estimate the beam-split-corrected THz channel $\mathbf{h}_k[m]$ and the physical DoAs $\vartheta_{k,l}$ for $m\in \mathcal{M}$ and $k\in \mathcal{K} = \{1,\cdots,K\}$ given the received pilot signals ${y}_k[m]$ in the presence of beam-split.

	\section{THz Channel Estimation}
	\textcolor{black}{Consider the downlink received signal model in (\ref{receivedSignal}). We assume a block-fading channel, wherein the coherence time is longer than the channel training time~\cite{channelEstThz2,ummimoTareqOverview}.}
	Let $\tilde{\mathbf{f}}_j[m]\in \mathbb{C}^{N_\mathrm{T}}$ be the beamformer vector, and the known pilot signals are $\tilde{\mathbf{S}}[m] = \text{diag}\{\tilde{{s}}_1[m],\dots,\tilde{{s}}_{J}[m]\} \in\mathbb{C}^{J\times J}$, where  ${J}$ is the number of pilot signals. Then, at the receiver side, each user collects the $ J\times 1$ received signals as
	\begin{align}
	\label{receivedPilots}
	\bar{\mathbf{y}}_k[m] = \tilde{\mathbf{S}}[m]\bar{\mathbf{F}}[m]\mathbf{h}_k[m] + \mathbf{n}_k[m],
	\end{align}
	where  $\bar{\mathbf{F}} = \tilde{\mathbf{F}}^\textsf{T}$, $ \tilde{\mathbf{F}}= [\tilde{\mathbf{f}}_1[m],\dots,\tilde{\mathbf{f}}_J[m]]^\textsf{T}\in \mathbb{C}^{J\times N_\mathrm{T}}$
	is beamformer matrix, $\bar{\mathbf{y}}_k = [y_k[1],\dots,y_k[M]]^\textsf{T}$ and ${\mathbf{n}}_k = [n_k[1],\dots,n_k[M]]^\textsf{T}$. Without loss of generality, selecting  $\bar{\mathbf{F}}[m] = \bar{\mathbf{F}}$ and $\tilde{\mathbf{S}}[m] = \mathbf{I}_J$, (\ref{receivedPilots}) is $\bar{\mathbf{y}}_k[m] = \bar{\mathbf{F}}{\mathbf{h}}_k[m] + \mathbf{n}_k[m]$, for which the least squares (LS) and MMSE solutions are  
	\begin{align}
	\mathbf{h}_k^{\mathrm{LS}}[m] &= (\bar{\mathbf{F}}^\textsf{H}\bar{\mathbf{F}})^{-1}\bar{\mathbf{F}}^\textsf{H}\bar{\mathbf{y}}_k[m]  ,  \\
	\mathbf{h}_k^{\mathrm{MMSE}}[m] &= \big(\mathbf{R}_k^{-1}[m] + \bar{\mathbf{F}}^\textsf{H} {\mathbf{R}}_k^{-1}[m]\bar{\mathbf{F}}  \big)^{-1}\bar{\mathbf{F}}^\textsf{H}\bar{\mathbf{y}}_k[m], 	\label{ls_MMSE}
	\end{align}
	respectively, where $\mathbf{R}_k[m] = \mathbb{E}\{\mathbf{h}_k [m]\mathbf{h}_k^{\textsf{H}}[m]\}$ is the channel covariance matrix. 
	
	\subsection{Sparse THz Channel Model}
	The main drawback of the conventional techniques in (\ref{ls_MMSE}) is the requirement of overdetermined system, i.e., $J\geq N_\mathrm{T}$, which results in high channel training overheads due to the UM number of antennas. However, the THz channel is extremely sparse in the angular domain (i.e., $L \ll N_\mathrm{T}$)~\cite{ummimoHBThzSVModel}. The angle domain representation of the channel is 
	\begin{align}
	\mathbf{x}_k[m] = \mathbf{F}^\textsf{H}\mathbf{h}_k[m] = \gamma \sum_{l = 1}^{L} \alpha_{k}^{m,l} \mathbf{u}_{k,m,l}e^{-j2\pi \tau_{k,l} f_m},
	\end{align}
	where $\mathbf{F}\in\mathbb{C}^{N_\mathrm{T}\times N}$ is an overcomplete discrete Fourier transform (DFT) matrix covering the whole angular domain with the resolution of $\rho = \frac{1}{2N}$.  The $n$th column of $\mathbf{F}$ is the steering vector $\mathbf{a}(\phi_n)\in \mathbb{C}^{N_\mathrm{T}}$ with $\phi_n = \frac{2n-N-1}{N}$ for $n = 1,\dots, N$. 
	$\mathbf{u}_{k,m,l}\in \mathbb{C}^{N}$ denotes the channel support vector corresponding to the angle domain representation with $\mathbf{u}_{k,m,l} = \mathbf{F}^\textsf{H}\mathbf{a}(\theta_{k,m,l}) = [\Sigma (\theta_{k,m,l} - \phi_1),\dots, \Sigma (\theta_{k,m,l} - \phi_N)]^\textsf{T}$, where $\Sigma (a) = \frac{\sin N \pi a/2}{\sin \pi a/2}$ is the Dirichlet sinc function. Due the power-focusing capability of $\Sigma(a)$, most of the power is focused on only small number of elements in the beamspace corresponding to the channel support~\cite{ummimoHBThzSVModel}. Therefore, $\mathbf{x}_k[m]$ is a sparse vector with sparsity level $L$, i.e., $||\mathbf{x}_k[m]||_0= L$, and the channel vector $\mathbf{h}_k[m]$ is approximated as $\hat{\mathbf{h}}_k[m]\approx \mathbf{F}\mathbf{x}_k[m]$.		Exploiting the sparsity of the THz channel, we consider the following underdetermined system, i.e.,
	\begin{align}
	\label{underdeterminedSystem}
	\mathbf{y}_k[m] =\mathbf{A}\mathbf{x}_k[m] + \mathbf{n}_k[m],
	\end{align}
	where $\mathbf{A} = \mathbf{BF}\in \mathbb{C}^{N_\mathrm{RF}\times N}$ and $\mathbf{B}\in \mathbb{C}^{N_\mathrm{RF}\times N_\mathrm{T}}$ is a frequency-independent matrix representing the precoder at the BS with $|[\mathbf{B}]_{i,j}| = \frac{1}{\sqrt{N_\mathrm{T}}}$.  	The sparsity of $\mathbf{x}_k[m]$ allows us to estimate $\mathbf{h}_k[m]$ with much less observations and pilot signals, i.e., $J = N_\mathrm{RF}\ll N_\mathrm{T}$.

	\begin{algorithm}[t]
		\begin{algorithmic}[1] 
			\caption{ \bf Beamspace Support Alignment}
			\Statex {\textbf{Input:}  $\mathbf{A}$, $\rho$, $\mathbf{y}_k[m]$, $f_m$ for $m\in \mathcal{M}$.} \label{alg:BSA}
			\Statex {\textbf{Output:} $\hat{\mathbf{h}}_k[m]$.}
			\State \textbf{for} $k \in\mathcal{K}$
			\State  Initialize:\hspace{-3pt} $l=1$,\hspace{-3pt} $\mathcal{I}_{l-1,k}[m]\hspace{-3pt} = \hspace{-3pt}\emptyset$,\hspace{-3pt} $\mathbf{r}_{l,k}[m] = \hspace{-3pt}\mathbf{y}_k[m]$\hspace{-3pt} for $m\in \mathcal{M}$.
			\State \textbf{for} $l\in \{1,\cdots, L\}$
			\State \indent\textbf{for} $m \in \mathcal{M}$
			\State \indent\indent $P_{k,n,l}[m]= | \mathbf{A}_n^\textsf{H}\mathbf{r}_{l,k}[m]|$  for $n = 1,\dots, N$ \label{step5}.
			\State \indent\indent ${n}_{k,m,l}^\star = \arg \maximize_n P_{k,n,l}[m]$.
			\State \indent\indent$\Delta_{k,m,l} = \lceil(1 - \frac{f_m}{f_c}) \frac{\phi_{{n}_{k,m,l}^\star}}{\rho}\rceil$.
			\State \indent\indent$\tilde{{P}}_{k,n,l}[m] = {P}_{k,n - \Delta_{k,m},l}[m]$. $\theta_{k,m,l} = \phi_{{n}_{k,m,l}^\star}$.
			\State \indent\textbf{end for}
			\State \indent $\bar{{P}}_{k,n,l} =  \sum_{{m}=1}^{M} \tilde{{P}}_{k,n,l}[{m}]$ for $n = 1,\dots, N$.
			\State \indent $ \bar{n}_{k,l}^\star= \arg \maximize_n \bar{{P}}_{k,n,l}$. $\vartheta_{k,l} = \phi_{\bar{n}_{k,l}^\star,l}$.
			\State \indent \textbf{for} $m \in \mathcal{M}$
			\State \indent\indent $\mathcal{I}_{l,k}[m] = \mathcal{I}_{l-1,k}[m] \bigcup \{\bar{n}_{k,l}^{\star} - \Delta_{k,m,l}\}$.
			\State \indent\indent $\mathbf{r}_{l+1,k}[m] \hspace{-3pt}= \hspace{-3pt} \mathbf{r}_{l,k}[m]\hspace{-3pt}- \hspace{-3pt}\mathbf{A}_{\mathcal{I}_{l+1,k}[m]} \mathbf{A}_{\mathcal{I}_{l+1,k}[m]}^\dagger\mathbf{y}_k[m]   $. 
			\State \indent  \textbf{end for}
			\State \textbf{end for} 
			\State  $\Xi_k[m] = \mathcal{I}_{1,k}[m]\bigcup \mathcal{I}_{2,k}[m], \dots, \bigcup\mathcal{I}_{L,k}[m]$. 
			\State $\mathbf{\hat{\mathbf{x}}}_k[m] = \mathbf{A}_{\Xi_k[m]}^\dagger \mathbf{y}_k[m]$. $\mathbf{\hat{\mathbf{h}}}_k[m] = \mathbf{F}\hat{\mathbf{x}}_k[m]$.
			\State \textbf{end for}
		\end{algorithmic} 
	\end{algorithm}
	
	\subsection{Beamspace Support Alignment}
	
	Due to the frequency-independent structure of $\mathbf{A}$, the spatial DoAs $\theta_{k,m,l}$ corresponding to the support of $\mathbf{x}_k[m]$ are deviated from  ${\vartheta_{k,l}}$ due to beam-split. Hence, the above sparse recovery techniques will not yield accurate results. In the proposed method, we align the deviated support of $\mathbf{x}_k[m]$ with respect to $m$ and generate a single beamspace spectrum for accurate channel estimation. 
	
	We start by determining the amount of index mismatch in the beamspace between the spatial and physical DoAs. Hence, we first compute the index ${n}_{k,m,l}^\star$ corresponding to the $l$th spatial channel DoA corresponding to the $k$th user and the $m$th subcarrier as $	{n}_{k,m,l}^\star = \arg \maximize_n P_{k,n,l}[m],$
	where $P_{k,n,l} [m] = | \mathbf{A}_n^\textsf{H}\mathbf{r}_{l,k}[m]|$ corresponds to the beamspace spectrum for the $m$th subcarrier. $\mathbf{r}_{l,k}[m]$ is the residual observation vector and equals to $ \mathbf{y}_{k}[m]$ for $l=1$, and $\mathbf{A}_n\in \mathbb{C}^{N_\mathrm{RF}}$ denotes the $n$th column of $\mathbf{A}$. Then, the $l$th spatial DoAs estimate  is $\hat{\theta}_{k,m,l} = \phi_{{n}_{k,m,l}^\star}$. {\color{black}Hence, the performance of the proposed BSA approach is limited by the angular resolution as error from the true physical DoA is $|\phi_{{n}_{k,m,l}^\star} - \theta_{k,m,l}|$.}  In order to find the physical DoAs, we first compute the index mismatch due to beam-split as
	\begin{align}
	\Delta_{k,m,l} = \lceil(1 - \frac{f_m}{f_c}) \frac{\phi_{{n}_{k,m,l}^\star}}{\rho}\rceil,
	\end{align}
	which measures the number of indices between the physical and spatial DoAs defined in (\ref{physical_spatial_directions}).  By shifting ${P}_{k,n,l}[m]$ by $\Delta_{k,m,l}$, we obtain the beam-split corrected spectrum as  $\tilde{{P}}_{k,n,l}[m] = {P}_{k,n - \Delta_{k,m,l},l}[m] $ which is the circularly shifted version of ${P}_{k,n,l}[m]$. In order to capture most of the power distributed  across all subcarriers due to beam-split and exploit multi-carrier channels, we incorporate the shifted beamspace spectrum for $m\in \mathcal{M}$ as  $	\bar{{P}}_{k,n,l} =  \sum_{{m}=1}^{M} \tilde{{P}}_{k,n,l}[{m}]$, which generates a single spectrum. 
	
	Next, we obtain the $l$th physical channel DoA $\vartheta_{k,l}$ as $\vartheta_{k,l} = \phi_{\bar{n}_{k,m}^\star}$, where $\bar{n}_{k,l}^\star= \arg \maximize_n \bar{{P}}_{k,n}$.  Combining the indices of DoAs in the set $\Xi_k[m]$ as $\Xi_k[m] = \bigcup_{l=1}^L \bar{n}_{k,l}^{\star} - \Delta_{k,m,l} $, the support vector is $	\hat{\mathbf{x}}_k [m] =  \mathbf{A}_{\Xi_k[m]}^\dagger \mathbf{y}_k[m],$
	for which,  we obtain the channel estimate as $\mathbf{\hat{\mathbf{h}}}_k[m] = \mathbf{F}\hat{\mathbf{x}}_k[m]$. The algorithmic steps of our proposed BSA approach are summarized in Algorithm~\ref{alg:BSA}. Using the channel and DoA estimates via BSA, we develop an FMTL scheme for THz channel estimation in the following.




	\section{Federated Multi-Task Learning}
	In FL, the users collaborate on training a learning model by computing the model parameters based on their local datasets~\cite{elbir_FL_PHY_Elbir2021Nov}. The proposed learning model has a single input while the output is comprised of the channel  and  DoA estimates. We design the input data $\mathcal{X}_k\in \mathbb{R}^{N_\mathrm{RF}\times 3}$ as	the combination of the real, imaginary and angle information of $\mathbf{y}_k[m]$. Thus, we have  $[\mathcal{X}_k]_1 = \Re\{\mathbf{y}_k[m]\}$, $[\mathcal{X}_k]_2 = \Im\{\mathbf{y}_k[m]\}$ and $[\mathcal{X}_k]_3 = \angle\{\mathbf{y}_k[m]\}$. Note that the angle information (i.e., $\angle\{\mathbf{y}_k[m]\}$) is additionally used to improve feature representation~\cite{elbir_FL_PHY_Elbir2021Nov}. The output  is represented as ${\mathcal{Y}}_k = \{\boldsymbol{\xi}_{1,k}, \boldsymbol{\xi}_{2,k} \}$, where the first term is comprised of the channel estimate $\hat{\mathbf{h}}_k[m]$ as  $\boldsymbol{\xi}_{1,k} = \left[\Re\{\hat{\mathbf{h}}_k[m]\}^\textsf{T},\Im\{\hat{\mathbf{h}}_k[m]\}^\textsf{T}\right]^\textsf{T}\in \mathbb{R}^{2N_\mathrm{T}}$. Further, the second term of $\mathcal{Y}_k$ is comprised of the $L$-sparse vector $\hat{\mathbf{x}}_k[m]$ as  $\boldsymbol{\xi}_{2,k} = |\hat{\mathbf{x}}_k^\textsf{T}[m]| \in \mathbb{R}^{N}$, for which the DoA angles corresponding to the non-zero entries yield the DoA estimates $\vartheta_{k,l}$. 	Let $\boldsymbol{\theta}\in \mathbb{R}^Q$ denote the set of learnable model parameters. Then, the trained model illustrated in Fig.~\ref{fig_Model} provides a nonlinear relationships $f_z(\boldsymbol{\theta})$ between the $i$th sample of the input and output as $\boldsymbol{\xi}_{z,k}^{(i)} = f_z(\boldsymbol{\theta}) \mathcal{X}_k^{(i)}$, where $f_z(\mathcal{X}_k^{(i)}| \boldsymbol{\theta})$ denotes model prediction given the input $\mathcal{X}_k^{(i)}$  for $z \in \{1,2\}$. Also, the $i$th sample of the local dataset is $\mathcal{D}_k^{(i)} = (\mathcal{X}_k^{(i)},\mathcal{Y}_k^{(i)})$.   Then, we can express the FMTL problem as
	\begin{align}
	\label{flTraining}
	\minimize_{\boldsymbol{\theta}} \hspace{3pt} \frac{1}{K}\sum_{k=1}^K \mathcal{L}_{\mathrm{Total},k}(\boldsymbol{\theta}) 
	\subjectto \hspace{2pt} f(\mathcal{X}_k^{(i)}| \boldsymbol{\theta}) = \mathcal{Y}_k^{(i)},
	\end{align}
	for $i = 1,\dots, \textsf{D}_k$, where $\textsf{D}_k = |\mathcal{D}_k|$ is the number of samples in the $k$th dataset for $k\in \mathcal{K}$. The total loss function in (\ref{flTraining}) is the sum of the losses for both tasks as $ \mathcal{L}_{\mathrm{Total},k}(\boldsymbol{\theta})  =  \sum_{z =  1}^{2}\omega_l\mathcal{L}_{z,k}(\boldsymbol{\theta})  $, where  $\mathcal{L}_{z,k} (\boldsymbol{\theta}) = \frac{1}{\textsf{D}_k} \sum_{i=1}^{\textsf{D}_k} ||f_z(\mathcal{X}_k^{(i)}| \boldsymbol{\theta}) - \mathcal{Y}_{z,k}^{(i)} ||_{\mathcal{F}}^2$ corresponds to the loss function at the $k$th user. To efficiently solve (\ref{flTraining}),  gradient descent is employed and the problem is solved iteratively, wherein model parameter update is performed at the $t$th iteration as $\boldsymbol{\theta}_{t+1} = \boldsymbol{\theta}_t - \kappa \frac{1}{K} \sum_{k=1}^{K}\boldsymbol{\beta}_k(\boldsymbol{\theta}_t)$ for $t = 1,\dots, T$. Here, $\kappa$ is the learning rate and $\boldsymbol{\beta}_k(\boldsymbol{\theta}_t)$ is the gradient vector as ${\boldsymbol{\beta}}_k(\boldsymbol{\theta}_t) = \nabla \mathcal{L}_k (\boldsymbol{\theta}_t + \Delta\boldsymbol{\theta}_t  )\in \mathbb{R}^{Q}$, where $\Delta\boldsymbol{\theta}_t$ denotes the noise terms added onto the model parameters due to wireless model transmission~\cite{elbir2020_FL_CE}. 

	\label{sec:complexity_overhead}
	{\color{black}	\textit{Model Training Overhead:} The communication overhead of both FMTL and CL can be given as the amount of data transmitted during training, i.e., $\mathcal{T}_\mathrm{FL} = 2 Q TK$, which includes the exchange of $Q$ model parameters for $T$ iterations, and $\mathcal{T}_\mathrm{CL} = \sum_{k = 1}^{K}\textsf{D}_kN_\mathrm{RF}$, where we assume the users transmit their $N_\mathrm{RF}$ received pilots to the BS, which performs labeling (channel and DoA estimation) process to obtain $\mathcal{Y}_k$ for $k\in \mathcal{K}$. 
	}
	
	\textit{Channel Training Overhead:} {\color{black}While the conventional techniques, e.g., LS and MMSE, require $J = N_\mathrm{T}$ orthogonal pilot signals as shown in (\ref{ls_MMSE}), the proposed learning-based approach does not rely on such requirements. Instead, the BSA approach employs much less channel training overhead with only $J= N_\mathrm{RF} $ pilot signals to accurately construct the sparse channel. }
	
	\textit{Complexity:} The computational complexity of the proposed BSA approach is mainly due to matrix multiplications in the steps 5 ($O(MN^2)$), 14 ($O(MN_\mathrm{RF}^2N^3)$) and 18 ($O(MN_\mathrm{RF}N + MN_\mathrm{T}N)$)  of Algorithm~\ref{alg:BSA}, respectively. Hence, the combined time complexity of the BSA approach is $O\big(MN(N(1+N_\mathrm{RF}^2N) + N_\mathrm{RF} + N_\mathrm{T})\big)$. 

	{\color{red}
		\begin{table}[h]
			\caption{Simulation Parameters
			}
			\label{tableSimulations}
			\centering
			\begin{tabular}{ccccccc}
				\hline 
				\hline	
				$f_c $ &$B$      &$M$    & $N_\mathrm{T}$ &$N_\mathrm{RF}$ & $L$ & $K$ \\
				\hline
				$300$ GHz& $15$ GHz & $128$ &$1024$ & $32$ & $5$ & $8$  \\
				\hline 
				$T$ & $V$      &$G$    & $\kappa$ &$\kappa_\mathrm{DCCN}$ & $\kappa_\mathrm{UDNN}$ & $\kappa_\mathrm{GAN}$\\
				\hline
				$100$& $1000$ & $500$ & $0.001$ & $0.05$ & $0.01$ & $0.005$ \\
				\hline
				\hline 
			\end{tabular}
		\end{table}
	}

	\section{Numerical Experiments}
	\label{sec:Sim}
	{\color{black}During simulations, the THz channel scenario is realized based on the parameters presented in Table~\ref{tableSimulations}. The user locations are selected as $\vartheta_{k,l}\in [-\frac{\pi}{2},\frac{\pi}{2}]$, and $\mathbf{F}$ is constructed for $N = 5N_\mathrm{T}$, and $[\mathbf{B}]_{i,j} = \frac{1}{\sqrt{N_\mathrm{T}}}e^{j{\varphi}}$, where $\varphi \sim \text{uniform}[-\frac{\pi}{2},\frac{\pi}{2}]$. }
	
	Data generation and training are handled over an hyperparameter optimization of the learning model in the FL toolbox in MATLAB on a PC with a 2304-core GPU. The proposed learning model  has total of $7$ convolutional layers and $2$ fully connected layers and $2$ dropout layers with $0.5$ probability as given in Fig.~\ref{fig_Model}. Hence, the total number of parameters is $Q = 1,196,928$~\cite{elbir2020_FL_CE}. The CNN model is trained for $T$ iterations with the learning rate $\kappa$ when $\omega_1 = .8$ and $\omega_2 = 0.2$. During data generation, we generated $V$ channel realizations, and we added AWGN on the input data for three signal-to-noise ratio (SNR) levels, i.e., $\text{SNR} = \{15,20,25\}$ dB for $G$ noisy realizations to provide robust performance~\cite{elbir_FL_PHY_Elbir2021Nov}. Also, the SNR for the noisy model transmission is $\mathrm{SNR}_\Delta = 20$ dB. Further, $80\%$ and $20\%$ of all generated data are chosen for training and validation datasets, respectively, which are beam-split-corrupted.  In order to make the local datasets non-i.i.d., we select the $k$th dataset DoA as $\vartheta_{k,l} \in [-\frac{\pi}{2} + \frac{\pi}{K} (k-1), -\frac{\pi}{2} + \frac{\pi}{K}k)$. The same dataset is used for all learning-based techniques, i.e., DCCN, UDNN and GAN.

	\begin{figure}[t]
		\centering
		{\includegraphics[draft=false,width=\columnwidth]{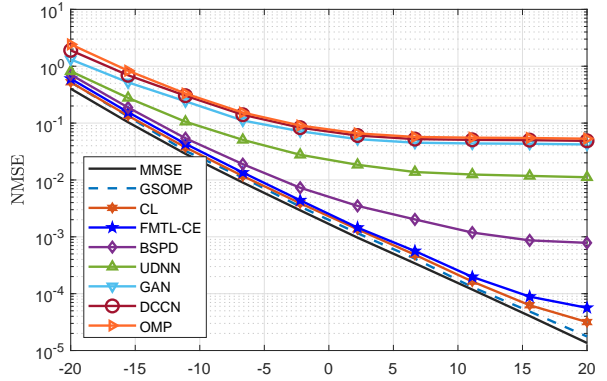} } 
		\caption{Channel estimation performance versus  SNR.
		}
		\vspace*{-5mm}
		\label{fig_NMSE_SNR}
	\end{figure}

	Fig.~\ref{fig_NMSE_SNR} shows the channel estimation performance of  our FMTL-BSA approach in comparison with the state-of-the-art techniques, e.g., MMSE, generalized simultaneous orthogonal matching pursuit (GSOMP)~\cite{channelEstThz2}, OMP, BSPD~\cite{channelEstThz_beamSplit}, DCCN~\cite{thzCE_CNN}, UDNN~\cite{thz_dl_Gao2022May} and GAN~\cite{thz_CE_GAN_Balevi2021Jan}, in terms of normalized MSE (NMSE), calculated as $\text{NMSE} = \frac{1}{J_\mathrm{T}KM}\sum_{i=1,k=1,m=1}^{J_\mathrm{T},K,M}\frac{|| \mathbf{h}_k[m]- \hat{\mathbf{h}}_k^{(i)}[m] ||_2^2 }{ || \mathbf{h}_k[m]||_2^2 }$, where $J_\mathrm{T}=500$ is the number of Monte Carlo trials.  Note that MMSE is selected as benchmark since it relies on the channel covariance matrix at each subcarrier, hence it presents beam-split-free performance.

	\begin{figure}[t]
		\centering
		{\includegraphics[draft=false,width=\columnwidth]{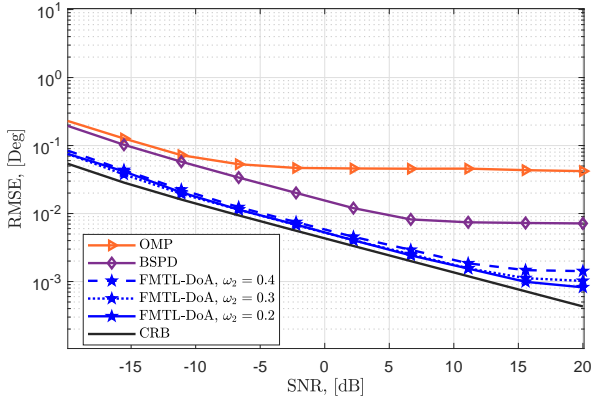} } 
		\caption{DoA estimation performance versus  SNR.
		}
		\label{fig_RMSE_SNR}
	\end{figure}

	In Fig.~\ref{fig_NMSE_SNR}, we observe that the proposed FMTL-CE approach achieves better NMSE as compared to the existing model-based (BSPD and OMP) and model-free (GAN, UDNN and DCCN) techniques, and it closely follows GSOMP and MMSE. There is a slight gap between FMTL-CE and CL, which stems from fact that CL has access to the whole training dataset while FMTL-CE is performed on local datasets.  Note that the GSOMP needs an additional hardware to generate virtual array response for $f_m\neq f_c$ for $m\in \mathcal{M}$, while the FMTL-CE does not involve such hardware requirement. The effectiveness of FMTL-CE is attributed to the mitigation of beam-split via support alignment with high resolution DoA estimates. The remaining methods, i.e. OMP, GAN, UDNN and DCCN, do not take into account the effect of beam-split for THz channel estimation. While the BSPD approach is designed to mitigate the beam-split, it fails to collects the accurate supports related to the aligned beamspace spectrum and suffers from poor physical DoA estimates due to the use of low resolution dictionary.

	Fig.~\ref{fig_RMSE_SNR} shows the DoA estimation performance of the proposed FMTL approach for $\omega_2 \in \{0.2,0.3,0.4\}$ when $\omega_1+\omega_2=1$. We can see the the proposed approach effectively estimates the user DoA angles as compared to the competing algorithms. We also observe a slight precision loss at high SNR regime due the precision loss which slightly changes as for different the regularization parameter $\omega_2$.

	\begin{figure}[t]
		\centering
		{\includegraphics[draft=false,width=\columnwidth]{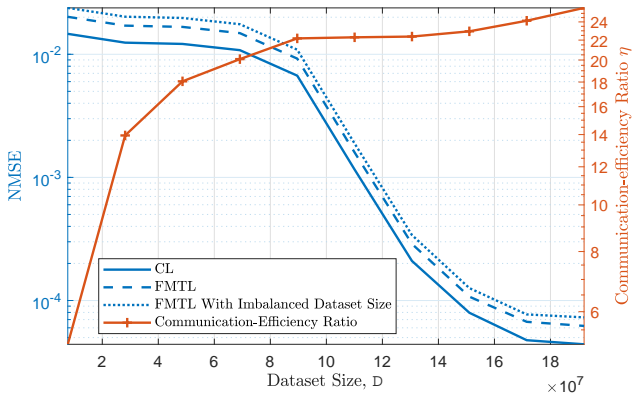} } 
		\caption{Comparison of CL and FMTL with respect to dataset size $\textsf{D}$.
		}
		\label{fig_NMSE_ETA_DATASET}
	\end{figure}

	Next, we investigate the NMSE performance and communication-efficiency ratio $\eta = \frac{\mathcal{T}_\mathrm{CL}}{\mathcal{T}_\mathrm{FL}}$ for CL and FMTL with respect to the size of the dataset as illustrated in Fig.~\ref{fig_NMSE_ETA_DATASET} for $\mathrm{SNR}=20$ dB. We can see that both CL and FMTL performs poor for small datasets, wherein the training is completed in fewer iterations. When large datasets are used to obtain improved channel estimation NMSE performance,  FMTL demonstrates approximately $25$ times lower communication overhead compared to that of CL. In such a scenario, the communication overhead of CL is computed as $\mathcal{T}_\mathrm{CL} = 8\cdot\textsf{D}_k\cdot 32 = 49.15\times 10^{9}	$, where the number of input-output tuples in the dataset is $\textsf{D}_k = 3M VG = 3\cdot 128\cdot 1000\cdot 500 = 192\times 10^{6} $. On the other hand, the communication overhead of FMTL is $\mathcal{T}_\mathrm{FL} = 2\cdot 1,196,928\cdot 100\cdot 8 = 1.91\times 10^{9}$, which exhibits approximately $25$ times lower than that of CL. We also present the performance of FMTL when the local datasets are imbalanced in Fig.~\ref{fig_NMSE_ETA_DATASET}, i.e., $\textsf{D}_k = \zeta_k\frac{\textsf{D}}{K}$, where $\zeta_k$ is selected such that $\zeta_k\in [0.7,1.3]$ and $\sum_{k=1}^{K}\zeta_k=1$. Evidently, imbalanced datasets yield poor NMSE performance since the aggregated learning models are unable to  represent the input-output relationship equally.


	\section{Conclusions}
	In this paper, we introduced an FMTL approach for THz channel and DoA estimation. The proposed approach is advantageous in terms of communication overhead and enjoys approximately $25$ ($32$) times lower model (channel) training overhead. Furthermore, we proposed BSA technique to align the deviated beamspace spectrum of  subcarriers and  mitigate the effect of beam-split without requiring additional hardware components.
	
	\bibliographystyle{IEEEtran}
	\bibliography{IEEEabrv,references_097}

	%
	%
	%
	%
	%
	%
	%
	%
	

	%

\end{document}